\documentclass[conference,10pt,a4paper]{IEEEtran}
\IEEEoverridecommandlockouts
\usepackage{amsmath}

\usepackage{amsopn}

\usepackage{float}

\usepackage{amssymb}

\usepackage{amsfonts}

\usepackage{amsthm}
\newtheorem{lemma}{Lemma}

\usepackage{graphicx}

\usepackage{booktabs}

\usepackage{caption3} 
\DeclareCaptionOption{parskip}[]{} 
\usepackage[small]{caption}

\usepackage{setspace,graphicx,multirow}
\usepackage{subcaption}

\usepackage{algorithm}
\usepackage[]{algpseudocode}
\makeatletter
\def\BState{\State\hskip-\ALG@thistlm}
\makeatother
\usepackage{algcompatible}

\newcommand*\diff{\mathop{}\!\mathrm{d}}

\usepackage{float}

\usepackage{geometry}
\usepackage{cite}

\usepackage{fancyhdr} 

\usepackage{color}
\ifodd 1
\newcommand{\com}[1]{\textbf{\color{blue} (COMMENT: #1)}} 
\else

\newcommand{\com}[1]{}
\fi

\usepackage{enumitem}
\setenumerate[1]{itemsep=0pt,partopsep=0pt,parsep=\parskip,topsep=0pt}
\setitemize[1]{itemsep=0pt,partopsep=0pt,parsep=\parskip,topsep=0pt}
\setdescription{itemsep=0pt,partopsep=0pt,parsep=\parskip,topsep=0pt}

\begin{document}
\bibliographystyle{IEEEtran}
\bstctlcite{IEEEexample:BSTcontrol}

\title{IRS-Aided Sectorized Base Station Design and \\ 3D Coverage Performance Analysis}

 \author{\IEEEauthorblockN{Xintong Chen,~\IEEEmembership{Student Member,~IEEE},
 Jiangbin Lyu,~\IEEEmembership{Member,~IEEE}, and
 Liqun Fu,~\IEEEmembership{Senior Member,~IEEE}}

\thanks{Manuscript submitted to IEEE IWQoS 2023 on 12 Feb. 2023; accepted 13 April 2023; published 27 July 2023. An associated Chinese patent was applied on 9 Aug. 2022 and granted on 1 Sep. 2023, under No. ZL202210948626.X. This work was supported in part by the Natural Science Foundation of Fujian 
	Province (No. 2023J01002), the Natural Science Fundation of Xiamen (No. 3502Z202372002), the Guangdong Basic and Applied Basic Research Foundation (No. 2023A1515030216), the National Natural Science Foundation of China (No. U23A20281, No. 61801408), and the Fundamental Research Funds for the Central Universities (No. 20720220078).}
\thanks{The authors are with the School of Informatics, Xiamen University, China, and also with the Shenzhen Research Institute of Xiamen University, China (email: xintongchen@stu.xmu.edu.cn; \{ljb, liqun\}@xmu.edu.cn). \textit{Corresponding author: Jiangbin Lyu}.}}

\maketitle



%
%
%
%
%
%
%

\begin{abstract}
Intelligent reflecting surface (IRS) is regarded as a revolutionary paradigm that can reconfigure the wireless propagation environment for enhancing the desired signal and/or weakening the interference, and thus improving the quality of service (QoS) for communication systems.
In this paper, we propose an IRS-aided sectorized BS design where the IRS is mounted in front of a transmitter (TX) and reflects/reconfigures signal towards the desired user equipment (UE).
Unlike prior works that address link-level analysis/optimization of IRS-aided systems, we focus on the system-level three-dimensional (3D) coverage performance in both single-/multiple-cell scenarios. 
To this end, a distance/angle-dependent 3D channel model is considered for UEs in the 3D space, as well as the non-isotropic TX beam pattern and IRS element radiation pattern (ERP), both of which affect the average channel power as well as the multi-path fading statistics.
Based on the above, a general formula of received signal power in our design is obtained, along with derived power scaling laws and upper/lower bounds on the mean signal/interference power under IRS passive beamforming or random scattering.
Numerical results validate our analysis and demonstrate that our proposed design outperforms the benchmark schemes with fixed BS antenna patterns or active 3D beamforming.
In particular, for aerial UEs that suffer from strong inter-cell interference, the IRS-aided BS design provides much better QoS in terms of the ergodic throughput performance compared with benchmarks, thanks to the IRS-inherent double pathloss effect that helps weaken the interference.
\end{abstract}


\section{Introduction}
With the commercial use of fifth-generation mobile networks, the demand for wide-area three-dimensional (3D) coverage and the explosive growth of wireless data continue to spur innovations in wireless communication technologies for higher spectral/energy efficiency and better quality of service (QoS).
Among others, some prominent technologies are proposed in the last decade including massive multiple-input multiple-output, ultra-dense network, and millimeter wave \cite{SmallCell5G}.
Although these technologies significantly enhanced the wireless network spectral efficiency, they also incurred increasingly \textit{more energy consumption} and \textit{higher hardware cost}, due to the deployment of more base stations (BSs) and/or relays in the network as well as mounting them with more active antennas requiring costly radio frequency (RF) chains, especially when operating at mmWave bands.
Moreover, the \textit{interference issue} caused by concurrent transmissions from more active BSs and/or active antennas is becoming the bottleneck for sustainable growth of network capacity \cite{PerformanceLimitsUDN}.
This is particularly the case for the 3GPP planned support of \textit{aerial users} \cite{3GPP36777}, which typically have a high probability of line-of-sight (LoS) channel with the ground BSs and thus suffer from strong inter-cell interference \cite{LyuNetworkUAV}.

To tackle the above challenges, intelligent reflecting surface (IRS), also known by other names such as reconfigurable intelligent surface (RIS), has recently emerged as a promising solution based on the new concept of smart and controlled wireless propagation environments \cite{IRSholographic,QingqingWuTutorial,RISsurvey2021,ZhengBeixiongSurvey2022}.
Specifically, IRS is a planar surface consisting of a massive number of low-cost passive reflecting elements that can be tuned dynamically to alter the amplitude and/or phase of the signal reflected by them, thus collaboratively reconfiguring the signal propagation to \textit{enhance the desired signal} and/or \textit{weaken the interference}, thereby improving the QoS of communications for both terrestrial and aerial user equipments (UEs) in the 3D space.
Compared with the conventional active relaying/beamforming, IRS does not require any active RF chain for signal transmission/reception but simply leverages passive wave reflection, thus leading to much lower hardware cost and energy consumption yet still operating spectral efficiently \cite{MDRenzoIRSvsRelay}.

The appealing advantages of IRS have attracted a great deal of interest in investigating IRS-aided wireless systems from various aspects and/or under different setups (see the recent surveys \cite{IRSholographic,QingqingWuTutorial,RISsurvey2021,ZhengBeixiongSurvey2022}), such as passive
beamforming design \cite{QQtwc,IRSchauYuen,IRSqqDiscrete,IRSshuowen}, 
IRS-aided orthogonal frequency division multiplexing
system \cite{IRSyifeiOFDM}\cite{IRSzhengBeixiong}, non-orthogonal multiple access \cite{IRSyangGangNOMA}\cite{IRSbeixiongNOMA}, physical layer security \cite{IRSsecureGuangchi}\cite{IRSsecureYCliang}, and so on.
The above works on IRS-aided wireless systems mainly aim to optimize the system performance at the link level with one or more IRSs serving one or few UEs at fixed locations, which show that the IRS-aided system can achieve significant spectral efficiency and/or energy efficiency improvement over the traditional system without IRS, with optimized IRS reflection coefficients.
However, such potential improvements are directly affected by the \textit{IRS electromagnetic characteristics} and its \textit{specific deployment} in the 3D environment, which are still in the early stage of study.


Regarding the electromagnetic nature of IRS and its associated channel characteristics, some practical experiments have been carried out on pioneering IRS prototypes \cite{TransShiJinPathlossmodeling, TransShiJinPathlossmodelingmmWave,IEEEAccessLinglongDaiIRSPrototype}.
First, based on antenna theory and practical measurements, the \textit{non-isotropic element radiation pattern (ERP)} of IRS is proposed in \cite{TransShiJinPathlossmodeling} and experimentally validated in \cite{TransShiJinPathlossmodelingmmWave}. The non-isotropic IRS ERP has a direct impact on the IRS-reflected channel power gains for UEs located at different \textit{angles} from the IRS, which, however, is typically ignored in the literature when considering IRS/UE(s) at fixed locations. 
Second, it is revealed that the signal experiences the \textit{double pathloss effect} when going through the two IRS-related links (i.e., the link between the transmitter (TX) and each IRS element, and the link between each IRS element and the UE), which depends on both the TX-IRS and IRS-UE \textit{distances}.
Third, in the ideal case with single LoS propagation path for the two IRS-related links, the authors in \cite{TransShiJinPathlossmodeling}\cite{TransShiJinPathlossmodelingmmWave} propose a general (single-path) formula for the UE's received signal power reflected through the IRS, which incorporates the IRS ERP and the free space pathloss model, and is verified by measurements in the electromagnetic chamber.
Nevertheless, for UEs distributed in a wide-area 3D space, there typically exists \textit{multi-path propagation} for the IRS-related links, which depends on the specific propagation environment and is distance/angle-dependent.
The above practical results and considerations motivate us to investigate the IRS-aided 3D coverage problem by incorporating the IRS ERP and wide-area 3D channel characteristics.

Due to the angle-dependent ERP and the double pathloss effect, the \textit{IRS deployment} needs careful design.
Given a total number of reflecting elements, there are various IRS deployment strategies by placing these elements at different locations, e.g., near the TX \cite{IEEEAccessLinglongDaiIRSPrototype,IRSArchitectureformmWave}, near the UEs \cite{LyuTWC2021} or both\cite{you2020deploy}, or dividing them into smaller-size IRSs that are distributed in the network\cite{JiangbinLyuSpatialThroughput,ShuowenZhangCapacityRegion,YunlongCaiIntelligentReflecting}.
In this paper, we focus on the near-TX deployment and propose a novel IRS-aided sectorized BS design for providing wide-area 3D coverage.
Note that different from the pioneering works (e.g., \cite{TransShiJinPathlossmodeling, TransShiJinPathlossmodelingmmWave,IEEEAccessLinglongDaiIRSPrototype,IRSArchitectureformmWave}) that focus on the IRS-aided communication prototype/architecture design and link-level performance analysis, we lay our basis on their works and focus on the \textit{system-level performance} for providing wide-area 3D coverage in both single- and multiple-cell scenarios.
Our main contributions are summarized as follows.
\begin{itemize}[leftmargin=0.14in]
\item First, we propose the \textit{IRS-aided sectorized BS design} where the IRS, as part of the BS, is mounted in front of a TX, providing coverage service in a sector/cell. We place the IRS in close proximity to the TX in order to alleviate the double pathloss effect, while highlighting the TX beam design, IRS ERP design, and the impact of their relative installation positions on the achievable antenna gain and beamwidth and hence the coverage range for spatially distributed UEs.
\item Second, the IRS ERP is incorporated with the distance/angle-dependent 3D channel models for both terrestrial and aerial UEs in a wide area. In particular, the impact of IRS ERP on the \textit{multi-path fading statistics} is derived, including the gains on the Rician factor and the mean fading power.
\item Third, 3D coverage performance analysis is performed based on our proposed design and the 3D channel model.
In particular, we analyze the impact of IRS passive beamforming/random scattering on the \textit{mean signal/interference power} by deriving their upper and lower bounds, based on which the double pathloss effect is manifested in explicit form and so is the power scaling law with the number of IRS elements. 
Interestingly, it is revealed that the double pathloss effect could be \textit{beneficial} in alleviating the inter-cell interference issue for the case with random scattering from non-serving IRSs in other cells, especially for UEs in the sky.
Based on these results, the signal-to-noise ratio (SNR) or signal-to-interference-plus-noise ratio (SINR) can be obtained for both ground and aerial UEs in the single- or multiple-cell scenarios.
Furthermore, the spatial distribution of the UEs' QoS in terms of ergodic throughput can be plotted as a \textit{3D coverage map} in the cell.
\end{itemize}

Finally, numerical results validate our analysis especially the impact of IRS ERP on the coverage performance, 
and demonstrate that our proposed design outperforms the benchmark schemes with fixed BS antenna pattern or active 3D beamforming under the same number of IRS/antenna elements. 
It is shown that our proposed design effectively addresses the uneven coverage issue due to sidelobe gaps in the fixed pattern scheme, and thus significantly improves the QoS in terms of both achievable throughput and the fairness among UEs. Moreover, for aerial UEs that suffer from strong inter-cell interference, the IRS-aided BS design provides much better QoS in terms of ergodic throughput compared with active 3D beamforming, thanks to the IRS-inherent double pathloss effect that helps weaken the interference.


The rest of this paper is organized as follows. The new IRS-aided sectorized BS design is presented in Section II. The 3D channel model and impact of IRS ERP are illustrated in Section III. Then, the 3D coverage performance analysis is provided in Section IV. Numerical results are shown in Section V. Finally, conclusions and future works are presented in Section VI.

\begin{table*}[!htbp]\small
\centering
\caption{Parameters and symbols for the IRS-aided sectorized BS design}
\begin{tabular}{l|l}
    \toprule
    Parameter/Symbol & Meaning \\
    \midrule
    $\boldsymbol{w}_{\textrm{tx}},\boldsymbol{w}_{\textrm{I}_n},\boldsymbol{w}_{u}$ & Positions of the TX, the $n$-th IRS element and UE $u$ \\
    $G_{\textrm{tx}},G_{\textrm{i}}$ & Maximum antenna power gain of the TX/IRS \\
    $F_{\textrm{tx}},F_{\textrm{i}}$ & Normalized power pattern of the TX/IRS \\
    $M_\textrm{Y},M_\textrm{Z}$ & Numbers of IRS elements along the Y-/Z-axis \\
    $d_\textrm{Y},d_\textrm{Z}$ & Element spacing along the Y-/Z-axis\\
    $\mu_\textrm{Y},\mu_\textrm{Z}$ & HPBW on the Y-/Z-planes \\
    $q_\textrm{Y},q_\textrm{Z}$ & Positive integers used to adjust the pattern shape \\
    $\alpha,\beta, S$ & The radius and the area of the elliptic footprint\\
    $D$ & TX-IRS distance \\
    $\theta,\phi$ & The elevation and azimuth angle in the spherical coordinate system \\
    \bottomrule
\end{tabular}
\end{table*}\label{ParametersForIRSaidedBSdesign}

\section{IRS-Aided Sectorized BS Design}\label{SectionDesign}

\begin{figure}
	\centering
	\includegraphics[width=0.75\linewidth,  trim=0 30 0 20,clip]{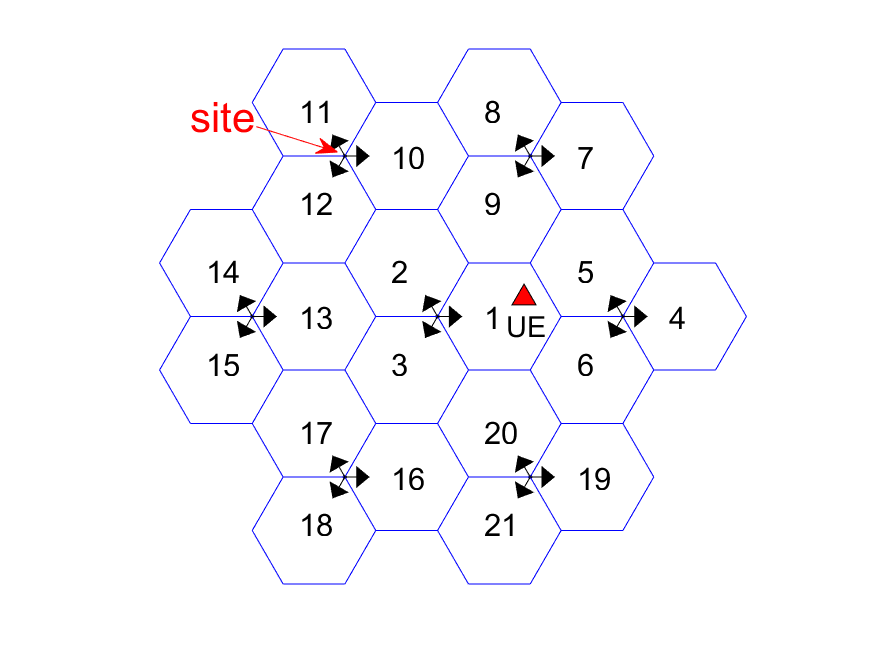}
	\caption{A cellular system with multiple sites, each with three sectors.\vspace{-2ex}}\label{IRSaidedBSinMultiCells}
\end{figure}

\begin{figure}
	\centering
	\includegraphics[width=0.85\linewidth,  trim=0 0 0 10,clip]{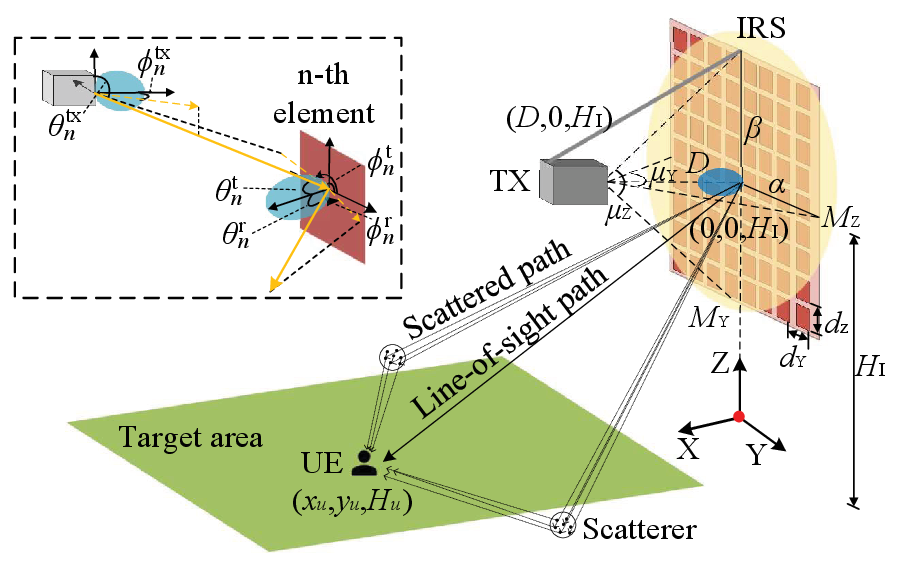}
	\caption{IRS-aided sectorized BS design in a single sector/cell.\vspace{-2ex}}\label{IRSaidedBS}
\end{figure}

Consider a cellular system with multiple sites as illustrated in Fig. \ref{IRSaidedBSinMultiCells}, where each site consists of three sectors/cells.
Different array configurations can be considered for each sector or cell, e.g., fixed pattern or 3D beamforming synthesized/enabled by 3GPP antenna ERP \cite{IEEEYongZengCellularUAV}, which are typically used in traditional cellular networks and hence taken as benchmark schemes.
In this section, we propose a novel IRS-aided sectorized BS design to serve UEs in each sector/cell, the key parameters of which are listed in Table \ref{ParametersForIRSaidedBSdesign}.

\subsection{Positions of TX, IRS and UE}
For each sector/cell, the IRS-aided sectorized BS consists of a single-antenna TX and a passive IRS with $N \triangleq M_{\textrm{Y}} \times M_{\textrm{Z}}$ reflecting elements arranged in the form of a uniform planar array (UPA), as illustrated in Fig. \ref{IRSaidedBS}.
Upon being illuminated by TX, the IRS can reconfigure the phase/amplitude of the signal reflected through each of its passive elements and beam towards the desired UE in the target area.
Take the red dot in Fig. \ref{IRSaidedBS} as the origin of a 3D Cartesian coordinate system $\mathcal{L}_{g}$.
Assume that the IRS is placed in the Y-Z plane with a height of $H_{\textrm{I}}$ meters (m), whose center position is given by $\boldsymbol{w}_{0} \triangleq (0,0,H_{\textrm{I}})$.
Here we use a flattened index $n\in\{1,\cdots,N\}\triangleq \mathcal{N}$ to replace the 2D index $(m_{\textrm{Y}},m_{\textrm{Z}})$ of an IRS element, with 
$n = (m_{\textrm{Y}}+\frac{M_{\textrm{Y}}}{2}-1)M_{\textrm{Z}} + (m_{\textrm{Z}}+\frac{M_{\textrm{Z}}}{2})$, $m_{\textrm{Y}} \in \left[ 1-\frac{M_{\textrm{Y}}}{2},\frac{M_{\textrm{Y}}}{2} \right]$, and $m_{\textrm{Z}} \in \left[1-\frac{M_{\textrm{Z}}}{2},\frac{M_{\textrm{Z}}}{2}\right]$.\footnote{For simplicity, we assume that both $M_{\textrm{Y}}$ and $M_{\textrm{Z}}$ are even numbers.}
Then, the $n$-th IRS element, denoted as $\textrm{I}_{n}$, locates at
\begin{equation}
 \boldsymbol{w}_{\textrm{I}_{n}} \triangleq \boldsymbol{w}_{0} + \boldsymbol{\overline{w}}_{\textrm{I}_{n}},
\end{equation}
where $\boldsymbol{\overline{w}}_{\textrm{I}_{n}} \triangleq \left(0,(m_{\textrm{Y}}-\frac{1}{2}) d_{\textrm{Y}}, (m_{\textrm{Z}}-\frac{1}{2}) d_{\textrm{Z}}\right)^{T}$; $d_{\textrm{Y}}$ and $d_{\textrm{Z}}$ denote the element spacing along the Y-axis and Z-axis, respectively, which are usually of subwavelength scale in the range between $\frac{\lambda}{10}$ and $\frac{\lambda}{2}$ \cite{TransShiJinPathlossmodeling}.
For simplicity, assume that the TX points towards the IRS center and locates at $\boldsymbol{w}_{\textrm{tx}} \triangleq (D, 0,H_{\textrm{I}})$, whereby the TX antenna boresight is perpendicular to the IRS surface.\footnote{The design can be readily extended to the non-perpendicular case.}
Besides, assume that the typical UE $u$ locates at $\boldsymbol{w}_{u} \triangleq \left( x_{u},y_{u},H_{u} \right)^{T}$, with $H_{u}$ denoting its height.

\subsection{TX Beam Design}\label{SectionTXbeam}
Assume that the TX is equipped with a directional antenna, while each UE has an isotropic one.
A general form of antenna power radiation pattern is given by $G_x F_x(\theta,\phi)$, with $G_x$ being the maximum antenna power gain, $F_x(\theta,\phi)$ being the normalized power pattern, $\theta$ and $\phi$ denoting the elevation and azimuth angles respectively in the local coordinate system\footnote{It is transformed by rotating the global coordinate system $\mathcal{L}_{g}$.} of the TX, as shown in Fig. \ref{IRSaidedBS}.
For elementary directional antennas such as horn or patch antennas, the cosine-shaped power radiation pattern is a good approximation \cite{antennaBook} that allows explicit trade-off between antenna beamwidth and directivity.
Therefore, for the purpose of exposition, assume that the TX power radiation pattern is given by $G_{\textrm{tx}} F_{\textrm{tx}}(\theta,\phi)$, with $F_{\textrm{tx}}(\theta,\phi)$ given in the form of
\begin{equation}\label{radiapattern_tx}
F_{\textrm{tx}}(\theta,\phi) = \sin^{2q_{\textrm{Z}}}
 \theta \cos^{2q_{\textrm{Y}}} \phi, \theta \in [0,\pi], \phi \in [-\pi/2,\pi/2],
\end{equation}
where $\theta^{\textrm{tx}}_{n},\phi^{\textrm{tx}}_{n}$ represents the elevation and azimuth angles,
and the positive integers $q_{\textrm{Y}}$ and $q_{\textrm{Z}}$ can be used to adjust the pattern shape.
The associated half power beamwidth (HPBW) on the azimuth and elevation planes, denoted by $\mu_{\textrm{Y}}$ and $\mu_{\textrm{Z}}$, respectively, can be obtained as
\begin{equation}\label{FormulaHPBWinY}
    \mu_{\textrm{Y}} = 2 \arccos 2^{-\frac{1}{2q_{\textrm{Y}}}},
\end{equation}
\begin{equation}\label{FormulaHPBWinZ}
    \mu_{\textrm{Z}} = 2 \arccos 2^{-\frac{1}{2q_{\textrm{Z}}}}.
\end{equation}

Note that when $q_{\textrm{Y}} \neq q_{\textrm{Z}}$, the HPBWs on the two principal planes are different. In this case, the perpendicular section of the pattern can be approximated by an ellipse, with which a typical rectangular IRS panel can be inscribed.
Therefore, for a given TX beam pattern and TX-IRS distance $D$, we can obtain the projection parameters, i.e., the radius $\alpha$ and $\beta$ of elliptic footprint as shown in Fig. \ref{IRSaidedBS}, which are given by $\alpha = D \tan \frac{\mu_{\textrm{Y}}}{2}$ and $\beta = D \tan \frac{\mu_{\textrm{Z}}}{2}$.
Then the area of the elliptic footprint can be expressed as $S \triangleq \pi \alpha \beta$.
To effectively utilize the IRS, we need to jointly design the TX beamwidth $\mu_{\textrm{Y}},\mu_{\textrm{Z}}$ and the TX-IRS distance $D$ so as to illuminate the whole IRS surface.
As a reference design, for the special case that the IRS panel is inscribed inside the elliptic footprint with the maximum rectangular area, we have the following relationships:
\begin{equation}
    M_{\textrm{Y}}d_{\textrm{Y}} = \sqrt{2}\alpha = \sqrt{2}D \tan \frac{\mu_\textrm{Y}}{2},
\end{equation}
\begin{equation}
    M_{\textrm{Z}}d_{\textrm{Z}} = \sqrt{2}\beta = \sqrt{2}D \tan \frac{\mu_\textrm{Z}}{2}.
\end{equation}

Another design consideration is the trade-off between antenna gain and installation convenience.
In general, the maximum antenna gain can be expressed as
\begin{equation}\label{MaxGain}
    G_{\textrm{tx}} = \frac{4\pi}{\int_{-\frac{\pi}{2}}^{\frac{\pi}{2}} \int_{0}^{\pi} F_{\textrm{tx}}(\theta,\phi) \sin \theta \diff \theta \diff \phi }.
\end{equation}
In particular, we have $G_{\textrm{tx}}=2(2q+1)$ when $q_{\textrm{Y}}=q_{\textrm{Z}}=q$.
Along with the HPBW formulas in \eqref{FormulaHPBWinY} and \eqref{FormulaHPBWinZ}, it implies the relationship between the TX antenna gain and its beamwidth, whereby a larger $G_{\textrm{tx}}$ corresponds to a smaller HPBW. 
Although narrower beamwidth is associated with higher antenna gain, it requires a longer distance $D$ to illuminate the IRS panel.
Therefore, there exists a trade off between the TX antenna gain $G_{\textrm{tx}}$ and TX-IRS distance $D$, both of which need to be properly chosen for cost reduction and installation convenience.

\subsection{IRS ERP Design}
We adopt the IRS ERP proposed in \cite{TransShiJinPathlossmodeling} and experimentally validated in \cite{TransShiJinPathlossmodelingmmWave}, which is given by $G_{\textrm{i}} F_{\textrm{i}}(\theta,\phi)$ with $G_{\textrm{i}}$ being the maximum power gain and $F_{\textrm{i}}(\theta,\phi)$ following the form of
\begin{equation}\label{IRSERPFormula}
    F_{\textrm{i}}(\theta,\phi) =
    \left\{
        \begin{array}{cc}
        (\cos \theta)^{\frac{G_{\textrm{i}}}{2}-1}, &  \theta \in [0,\frac{\pi}{2}], \phi \in [0,2\pi], \\
        0, &  \theta \in (\frac{\pi}{2},\pi], \phi \in [0,2\pi].
        \end{array}
    \right.
\end{equation}
For an IRS element, its ERP affects both the incident and reflected signals as illustrated in Fig. \ref{IRSaidedBS}, where $\theta^{x}_{n}$ and $\phi^{x}_{n}$ represent the elevation and azimuth angles respectively in the local coordinate system of the IRS, with $x\in \{\textrm{t,r}\}$ representing the incident and reflected signal at $\textrm{I}_{n}$.

Similar to the impact of TX beam pattern on the transmitted signal, the IRS ERP explicitly characterizes the 3D angle dependence of the signal incident/reflected by each IRS element.
In particular, the choice of a maximum power gain $G_{\textrm{i}}$ is a design factor that tradeoffs between radiation beamwidth and directivity gain, which in turn affects the 3D coverage performance for UEs distributed in a wide area/space.
Finally, besides ERP, the joint passive beamforming/random scattering by all IRS elements also plays a key role in affecting the signal/interference power and hence the overall coverage performance, as will be discussed later in Section \ref{SectionAnalysis}.

\section{3D Channel Model and Impact of IRS ERP}\label{SectionChannelModel}
In this section, we introduce the distance/angle-dependent 3D channel model in the considered area, with emphasis on the impact of IRS ERP on the channel statistics.
For a typical cell/sector, consider the downlink communication from the BS to its served UEs, denoted by $\mathcal{U} \triangleq \{1,\cdots, U\}$,
whereas the results obtained can be similarly applied to the uplink communication as well. 
To focus on the coverage performance, assume that the served UEs are assigned with orthogonal-time Resource Blocks (RBs), i.e., time division multiple access (TDMA) or time-sharing is adopted.\footnote{TDMA is in general superior over FDMA due to hardware limitation of IRS passive reflection, which can be made time-selective, but not frequency-selective \cite{QQirsMag}. Other multiple access schemes are left for extended future work.}
For a typical UE $u\in\mathcal{U}$ assigned on a typical RB with bandwidth $B$ and transmit power $P_{\textrm{t}}$, we introduce the IRS-related channels in the following.\footnote{Based on our design in Section \ref{SectionDesign}, the TX beam points towards the IRS and hence the direct TX-UE link is assumed to be negligible for simplicity.}

\subsection{TX-IRS Channel}
Due to the short distance, we assume that the whole IRS panel is in the near-field of the TX.
As a result, the angle and distance between the TX and each IRS element $\textrm{I}_{n}$, $n\in\mathcal{N}$, need to be considered individually.
Furthermore, assume free space propagation between the TX and $\textrm{I}_n$. 
When considering isotropic radiation patterns at both TX and IRS sides, the TX-$\textrm{I}_{n}$ channel amplitude can be expressed as
\begin{equation}
    \lvert h_{\textrm{TI}_{n}} \rvert \triangleq \sqrt{ g_{\textrm{TI}_{n}} } = \sqrt{ \beta_{\textrm{0}}(d_{\textrm{TI}_{n}})^{-2} },
\end{equation}
where $g_{\textrm{TI}_{n}}$ denotes the channel power gain, $d_{\textrm{TI}_{n}}$ denotes the TX-$\textrm{I}_{n}$ distance, and $\beta_{\textrm{0}} = (\frac{4\pi f_{\textrm{c}}}{c})^{-2}$ denotes the channel power gain at a reference distance of 1 m, with $f_{\textrm{c}}$ denoting the carrier frequency and $c$ denoting the speed of light.

When \textit{non-isotropic} radiation patterns are applied at the TX and IRS, the TX-$\textrm{I}_{n}$ channel amplitude is given by
\begin{equation}\label{EQhtiprime}
    \lvert h^{\prime}_{\textrm{TI}_{n}}\rvert \triangleq
    \sqrt{ G_{\textrm{tx}} G_{\textrm{i}} F_{\textrm{tx}}(\theta^{\textrm{tx}}_{n},\phi^{\textrm{tx}}_{n}) F_{\textrm{i}}(\theta^{\textrm{t}}_{n},\phi^{\textrm{t}}_{n}) g_{\textrm{TI}_{n}} }.
\end{equation}


\subsection{IRS-UE Channel and Impact of IRS ERP}\label{SectionImpactERP} 
\subsubsection{Distance/angle-dependent channel model}
Assume far-field propagation from the IRS to the typical UE $u$, and hence the distance $d_{\textrm{I}_{n} u}$, direct (i.e., LoS direction) elevation and azimuth angles $(\theta^{\textrm{r}}_{n},\phi^{\textrm{r}}_{n})$ from each IRS element $\textrm{I}_{n}$ to the UE are approximately treated as the same for all elements $n\in\mathcal{N}$, denoted as $d_u$, $\theta_u$ and $\phi_u$, respectively. Specifically, the distance from the IRS center $\boldsymbol{w}_{0}$ to the UE position $\boldsymbol{w}_{u}$ is given by
\begin{equation}
    d_u\triangleq \sqrt{x_u^2+y_u^2+(H_u-H_{\textrm{I}})^2},
\end{equation}
while the LoS direction of the UE, denoted as $\boldsymbol{\Omega}_u\triangleq (\theta_u,\phi_u)$, with reference to the IRS's boresight direction, is given by
\begin{equation}
    \theta_u \triangleq \arccos\frac{x_u}{d_u},
\end{equation}
\begin{equation}
    \phi_u \triangleq \arccos \frac{y_{u}}{\sqrt{y_u^2+(H_u-H_{\textrm{I}})^2}}.
\end{equation}

Therefore, when considering isotropic radiation patterns at both IRS/UE sides, the $\textrm{I}_{n}$-UE $u$ channel amplitude is given by
\begin{equation}\label{IRStoUEchannelformula}
    \lvert h_{\textrm{I}_{n}u} \rvert \triangleq \sqrt{g_u} \xi_{n},
\end{equation}
where the average channel power gain $g_u=\textrm{PL}_u^{-1}$ is defined as the reciprocal of the large-scale pathloss $\textrm{PL}_u$, and the term $\xi_{n}$ accounts for small-scale fading.
In general, the pathloss $\textrm{PL}_u$ between the BS (or IRS) and a UE $u$ in the 3D space is a function of the BS-UE distance and the UE's elevation angle, whose function form depends on the environment and the specific LoS/NLoS channel condition.
For the purpose of exposition, we adopt the LoS/NLoS pathloss functions with their associated probabilities as specified by 3GPP in \cite{3GPP36777} for the urban macro (UMa) scenario.

On the other hand, the term $\xi_{n}$ can be modeled as a random variable (RV) that characterizes the multi-path fading effect.
For simplicity, we consider the case with half-wavelength element spacing $d_\textrm{Y}=d_\textrm{Z}=\lambda/2$ and hence assume that the fading terms $\xi_{n}$ of all IRS elements $n\in\mathcal{N}$ are independent and identically distributed (i.i.d.), which follow the Rician distribution $\textrm{Rice}(\upsilon,\sigma)$ with scale parameters $\upsilon = \sqrt{\frac{K}{K+1}}$ and $\sigma = \sqrt{\frac{1}{2(K+1)}}$ and hence a mean power of $\mathbb{E}[(\xi_{n}) ^2]=1$.
In particular, the Rician K factor represents the ratio of mean power of the direct LoS path against that of all other NLoS paths, which is distance- and/or angle-dependent in a given propagation environment, and also depends on the LoS/NLoS pathloss condition.
In the case of NLoS pathloss condition, we assume $K=0$ which reduces to Rayleigh fading.
In the case of LoS pathloss condition, we choose a distance/angle-dependent Rician K factor for ground/aerial UEs, respectively.
For ground UEs, based on \cite{3GPPTR25996}, we have
\begin{equation}
    K=13-0.03 d_{u}\quad (\textrm{dB}),
\end{equation}
which decreases with the BS-UE distance $d_u$ (m).
For aerial UEs above the BS height, based on \cite{TWCChangshengYou3DRicianFadingUAV}, the angle-dependent Rician K factor can be modeled as
\begin{equation}
    K = A_{1}\textrm{exp}(A_{2}\theta_u^{\prime}),
\end{equation}
where $A_{1}$ and $A_{2}$ are constant coefficients determined by the environment, and $\theta_u^{\prime}\triangleq \arcsin\frac{\lvert H_u-H_{\textrm{I}} \rvert}{d_u}$ is the elevation angle of the UE with reference to the horizontal plane at height $H_\textrm{I}$ that the IRS center resides.
Then we have $K_{\textrm{min}} \leq K \leq K_{\textrm{max}}$, where $K_{\textrm{min}} = A_{1}$ and $K_{\textrm{max}} = A_{1}e^{A_{2}\pi/2}$.

In addition to the fading amplitude $\xi_n$ and the LoS direction $\Omega_u$, we further consider the direction of departure (DoD),\footnote{Due to the assumed isotropic antenna pattern of the UE, the direction of arrival (DoA) at the UE is not considered for simplicity.} denoted by $\boldsymbol{\Omega}_n^{\textrm{r}}\triangleq (\theta_n^{\textrm{r}},\phi_n^{\textrm{r}})$, of the reflected NLoS paths from an IRS element $\textrm{I}_{n}$ to UE $u$, which is similarly assumed to follow the same distribution for all IRS elements $n\in\mathcal{N}$.
Specifically, define the normalized power angular spectrum of the NLoS reflected paths as $P_{\textrm{S}}( \boldsymbol{\Omega}^{\textrm{r}})$, which then integrates to a total power of $1/(K+1)$, i.e.,
\begin{equation}
    \oint P_{\textrm{S}}(\boldsymbol{\Omega}^{\textrm{r}})\diff\boldsymbol{\Omega}^{\textrm{r}}= \frac{1}{K+1}.
\end{equation}

\subsubsection{Impact of IRS ERP on multi-path fading statistics}\label{subsubSectionERPonMultipath}
When the non-isotropic radiation pattern is considered, i.e., by incorporating the IRS ERP, the multi-path fading statistics of the $\textrm{I}_{n}$-UE $u$ channel will be affected, including the Rician K factor and the mean fading power. Therefore, we have the following lemma to describe the impact of IRS ERP explicitly.
\begin{lemma}
Define the gains of the Rician K factor and the mean fading power as
\begin{equation}
    G_{K} \triangleq \frac{K^{\prime}}{K},
\end{equation}
\begin{equation}\label{rhoDefinition}
    \rho \triangleq \frac{\mathbb{E}[(\xi_{n}^{\prime})^2]}{\mathbb{E}[(\xi_{n}) ^2]} =\mathbb{E}[(\xi_{n}^{\prime}) ^2],
\end{equation}
respectively, where $\xi_{n}^{\prime}$ denotes the new fading term with the impact of IRS ERP, which can be shown to follow the Rician distribution $\textrm{Rice}(\upsilon^{\prime},\sigma^{\prime})$ with a mean power of $\rho$ and scale parameters $\upsilon^{\prime} =\sqrt{\rho}\upsilon= \sqrt{\frac{\rho K^{\prime}}{K^{\prime}+1}}$ and $\sigma^{\prime} =\sqrt{\rho}\sigma= \sqrt{\frac{\rho}{2(K^{\prime}+1)}}$.
\end{lemma}

\emph{Proof:} See Appendix \ref{AppenIRSERPonChannel}.\hfill $\blacksquare$

As a result, with the impact of IRS ERP, the $\textrm{I}_{n}$-UE $u$ channel amplitude is rewritten as
\begin{equation}\label{hprimtheory}
    \lvert h^{\prime}_{\textrm{I}_{n}u} \rvert \triangleq \sqrt{g_u} \xi^{\prime}_{n}.
\end{equation}
Based on similar analysis in \cite{TransHaibingYangImpactDirectionalAntenna} for directional antennas, we derive $G_K$ and $\rho$ in the following.
First, for the (potentially present) LoS component (i.e., $K\neq 0$), its power $K/(K+1)$ is scaled by the IRS pattern gain $G_{\textrm{i}}F_{\textrm{i}}(\boldsymbol{\Omega}_u)$ along the LoS direction $\boldsymbol{\Omega}_u\triangleq (\theta_u,\phi_u)$. 
Second, for the NLoS components, its mean power $E_\textrm{NLoS}$ can be obtained by weighing the power angular spectrum $P_{\textrm{S}}(\boldsymbol{\Omega}^{\textrm{r}})$ with the ERP $G_{\textrm{i}}F_{\textrm{i}}(\boldsymbol{\Omega}^{\textrm{r}})$, i.e.,
\begin{equation}\label{NLOSmeanpowerFormula}
    E_\textrm{NLoS} = G_{\textrm{i}} \oint P_{\textrm{S}}(\boldsymbol{\Omega}^{\textrm{r}})F_{\textrm{i}}(\boldsymbol{\Omega}^{\textrm{r}})\diff\boldsymbol{\Omega}^{\textrm{r}}.
\end{equation}
Therefore, by the definition of Rician K factor, we have
\begin{equation}
    K'= \frac{G_{\textrm{i}}F_{\textrm{i}}(\boldsymbol{\Omega}_u) \cdot K/(K+1)}{E_{\textrm{NLoS}}}=\frac{G_{\textrm{i}}F_{\textrm{i}}(\boldsymbol{\Omega}_u) \cdot K}{(K+1) E_{\textrm{NLoS}}},
\end{equation}
\begin{equation}
    \rho =E_{\textrm{NLoS}}+ G_{\textrm{i}}F_{\textrm{i}}(\boldsymbol{\Omega}_u) \cdot K/(K+1),
\end{equation}
and hence $G_K=\frac{G_{\textrm{i}}F_{\textrm{i}}(\boldsymbol{\Omega}_u) }{(K+1) E_{\textrm{NLoS}}}$.

\begin{figure}
	\centering
	\includegraphics[width=0.65\linewidth,  trim=0 0 0 0,clip]{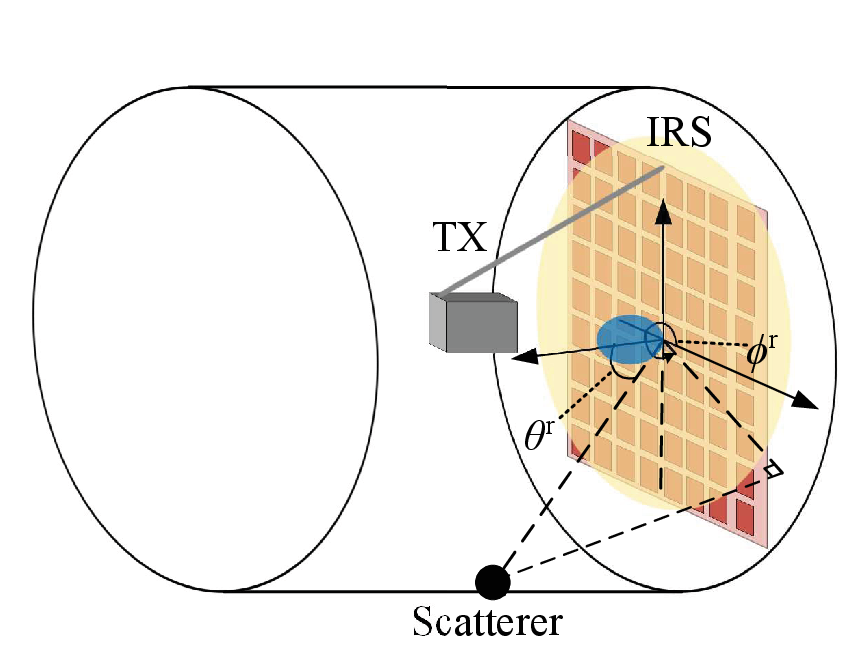}
	\caption{A 3D cylinder model for scatterers.\vspace{-2ex}}\label{3DcylinderScatter}
\end{figure}

For illustration purpose, we consider an example of power angular spectrum for $P_{\textrm{S}}(\boldsymbol{\Omega}^{\textrm{r}})$ and model the scatterers as non-uniformly distributed on the surface of a cylinder centered at $\textrm{I}_{0}$, as shown in Fig.\ref{3DcylinderScatter}.
Specifically, $P_{\textrm{S}}(\boldsymbol{\Omega}^{\textrm{r}})$ is given by
\begin{equation}\label{powerangularspectrum}
    P_{\textrm{S}}(\boldsymbol{\Omega}^{\textrm{r}}) = 
    \left\{
        \begin{array}{cc}
        \frac{1}{K+1}f_1(\theta^{\textrm{r}})f_2(\phi^{\textrm{r}}), & \boldsymbol{\Omega}^{\textrm{r}} \in  \boldsymbol{U}, \\
        0, &\textrm{ otherwise,}
        \end{array}
    \right.
\end{equation}
where $\boldsymbol{U}\triangleq \left\{(\theta,\phi) \big| \theta \in \left[ \theta^{\textrm{L}},\theta^{\textrm{H}} \right], \phi \in \left[ \phi^{\textrm{L}},\phi^{\textrm{H}} \right] \right\}$ defines the angular ranges, while $f_1(\theta^{\textrm{r}})$ and $f_2(\phi^{\textrm{r}})$ follow the cosine probability density function (PDF) and the normalized von Mises PDF \cite{WCNCLinzhouZengScatter}, respectively, which can be expressed\footnote{Note that the von Mises PDF in \cite{WCNCLinzhouZengScatter} is truncated and normalized in order to model scatterers in a certain angular range.} as
\begin{equation}\label{PDFEleDistribution}
\begin{aligned}
    f_1(\theta) \triangleq \frac{\pi}{4 \theta_{\textrm{m}}} \cos \left(\frac{\pi}{2} \frac{\theta-\theta_{\textrm{g}}}{\theta_{\textrm{m}}} \right), \left| \theta-\theta_{\textrm{g}} \right| \leq \theta_{\textrm{m}} \leq \frac{\pi}{2},
\end{aligned}    
\end{equation}
\begin{equation}
\begin{aligned}
    f_2(\phi) \triangleq \frac{f_0(\phi)}{P}, \phi^{\textrm{L}} \leq \phi \leq \phi^{\textrm{H}},
\end{aligned}
\end{equation}
where $\theta \triangleq \frac{\pi}{2}-\theta^{\textrm{r}}$, $f_{0}(\phi) \triangleq \frac{e^{k \cos( \phi-\phi_\textrm{g} ) }} {2\pi I_0(k)}$ and $P \triangleq \int_{\phi^{\textrm{L}}}^{\phi^{\textrm{H}}} f_0(\phi) d\phi$;
$\theta_{\textrm{g}}$ and $\theta_{\textrm{m}}$ are related to the mean angle and variance, respectively; $I_0(\cdot)$ is the zeroth-order modified Bessel function of the first kind; $\phi_{\textrm{g}}$ is the mean angle at which the scatterers are distributed in the IRS plane; and $k$ controls the spread around the mean angle.

Note that the specific angular distribution of scatterers has a direct impact on the multi-path fading statistics due to the angle-dependent IRS ERP. Specifically, in this case, different elevation angular distributions $f_1(\theta)$ by varying $\theta_{\textrm{g}}$ and $\theta_{\textrm{m}}$ will affect the NLoS mean power $E_{\textrm{NLoS}}$ and hence the $\textrm{I}_n$-UE $u$ channel statistics, as will be illustrated later in Section \ref{RicianProveMC}.

\subsection{TX-IRS-UE Channel}
Based on the above model and analysis, 
by considering non-isotropic radiation patterns at the TX and IRS, the cascaded TX-$\textrm{I}_{n}$-UE $u$ channel can be written as
\begin{equation}
    h_{\textrm{TI}_{n}u} \triangleq h^{\prime}_{\textrm{TI}_{n}} \Gamma_{n} h^{\prime}_{\textrm{I}_{n}u}
    = \lvert h^{\prime}_{\textrm{TI}_{n}} \rvert 
    \lvert h^{\prime}_{\textrm{I}_{n}u} \rvert A_{n}e^{j(\zeta_{n}+\psi_n)},
\end{equation}
where $\Gamma_{n}\triangleq A_{n}e^{j\zeta_{n}}$ denotes the reflection coefficient of $\textrm{I}_{n}$ with the amplitude $A_{n} \in [0,1]$ and the phase $\zeta_{n} \in [0,2\pi)$, and $\psi_n\triangleq\angle(h^{\prime}_{\textrm{TI}_{n}}h^{\prime}_{\textrm{I}_{n}u})$ denotes the cascaded channel phase, which can be estimated by existing methods such as \cite{IRSzhengBeixiong}.

As a result, the overall TX-IRS-UE channel is given by
\begin{equation}
    h_{\textrm{TI}u} \triangleq \sum\nolimits_{n\in\mathcal{N}} h_{\textrm{TI}_{n}u}.
\end{equation}
Finally, the received power by UE $u$ is given by
\begin{align}
    P_{\textrm{r},u} &\triangleq P_{\textrm{t}} \left| h_{\textrm{TI}u} \right| ^{2}\notag\\
    &=P_{\textrm{t}} G_{\textrm{tx}} G_{\textrm{i}} g_u 
    \Big| \sum\limits_{n\in\mathcal{N}}  \sqrt{ F_{\textrm{combine}} g_{\textrm{TI}_n} } \xi^{\prime}_n A_n e^{j(\zeta_n+\psi_n)} \Big|^{2},\label{Pru}
\end{align}
where $F_{\textrm{combine}} \triangleq F_{\textrm{tx}}(\theta^{\textrm{tx}}_{n},\phi^{\textrm{tx}}_{n}) F_{\textrm{i}}(\theta^{\textrm{t}}_{n},\phi^{\textrm{t}}_{n})$.

\section{3D Coverage Performance Analysis}\label{SectionAnalysis}
A general formula is obtained in \eqref{Pru} for the received power of a typical UE $u$ in our IRS-aided sectorized BS design, which is applicable to the desired signal from the serving IRS, as well as the interference from non-serving IRSs in other cells.
In this section, we first analyze the impact of IRS passive beamforming/random scattering on the signal/interference power, respectively, and then discuss the SNR/SINR and ergodic throughput in single-/multiple-cell scenarios.

\subsection{Impact of IRS Passive Beamforming/Random Scattering on Signal/Interference Power} \label{SectionIRSpassiveBeamRandomScatter}
Based on \eqref{Pru}, the UE's received power depends on the 3D channel model and the directional TX pattern and IRS ERP, as well as the IRS phase shifts that could lead to different power scaling laws.
In the following, we derive expressions for the signal or interference power as well as upper/lower bounds on their mean values, under IRS passive beamforming or random scattering, respectively.

\subsubsection{IRS passive beamforming}
In this case, we assume continuous phase shift capability\footnote{The case with discrete phase shift \cite{IRSqqDiscrete} is left for our extended future work.} at the IRS which can adjust the phase shift $\zeta_{n}$ such that the $N$ reflected signals are of the same phase at UE $u$ by setting $\zeta_{n}=-\psi_n=-\angle(h^{\prime}_{\textrm{TI}_{n}}h^{\prime}_{\textrm{I}_{n}u})$.
In addition, assume for simplicity that $A_{n} = A$, $n\in\mathcal{N}$. As a result, under the above passive beamforming by the serving IRS, the received power of UE $u$ is given by
\begin{equation}
    P_{\textrm{r},u}=P_{\textrm{t}} G_{\textrm{tx}} G_{\textrm{i}} g_u A^2 
    \left| \sum\nolimits_{n\in\mathcal{N}} \sqrt{ F_{\textrm{combine}} g_{\textrm{TI}_n} } \xi^{\prime}_n \right|^{2}. \label{Prphasealign}
\end{equation}

Define $b_{\textrm{0}}\triangleq P_{\textrm{t}} G_{\textrm{tx}} G_{\textrm{i}} g_u A^2$ and $w_{n}\triangleq \sqrt{ F_{\textrm{combine}} g_{\textrm{TI}_{n}}}$ for convenience of derivations. The coefficient $b_0$ depends on $g_u$ and hence the IRS-UE pathloss $\textrm{PL}_u$. The weight $w_{n}$ depends on $F_{\textrm{combine}}$ and $g_{\textrm{TI}_n}$, which in turn depends on the TX antenna/IRS ERP pattern as well as a second pathloss between the TX and IRS.
Therefore, it can be seen from \eqref{Prphasealign} that the UE's received power depends on both the TX-IRS pathloss and the IRS-UE pathloss, and hence suffers from the \textit{double pathloss} effect.

Fortunately, passive beamforming can be a rescue with sufficiently large number $N$ of IRS elements.
In the following, we derive the upper or lower bound of $P_{\textrm{r},u}$ and reveal the power scaling law with $N$.
When $w_{n}$ takes the maximum (minimum) value\footnote{The maximum (minimum) $w_{n}$ typically happens at the IRS center (edge).} for all $n\in\mathcal{N}$, denoted as $w_{\textrm{max}}$ ($w_{\textrm{min}}$), we have
\begin{equation}
    b_{0}w^2_{\textrm{min}}\big| \sum\nolimits_{n\in\mathcal{N}} \xi^{\prime}_n \big|^{2} \leq P_{\textrm{r},u} \leq b_{0}w^2_{\textrm{max}}\big| \sum\nolimits_{n\in\mathcal{N}} \xi^{\prime}_n \big|^{2}.
\end{equation}
Denote $\Xi \triangleq \sum\nolimits_{n\in\mathcal{N}} \xi^{\prime}_{n}$. For i.i.d. $\xi^{\prime}_{n}$, $n\in\mathcal{N}$, by the central limit theorem (CLT), $\Xi$ can be approximated by the Gaussian distribution for practically large $N$ \cite{LyuTWC2021}, i.e.,
\begin{equation}
\begin{aligned}
     \Xi \triangleq \sum\nolimits_{n\in\mathcal{N}} \xi^{\prime}_{n} \stackrel{\textrm{approx.}}\sim \textrm{Gaussian} \left( N\mathbb{E}\{  \xi^{\prime}_{n} \}, N \textrm{var}\{  \xi^{\prime}_{n} \} \right),
\end{aligned}
\end{equation}
where the mean and variance of $\xi^{\prime}_{n}$ are respectively given by
\begin{equation}\label{ExpectRicianDis}
    \mathbb{E}\{ \xi^{\prime}_{n} \} \triangleq \sqrt{\frac{\rho\pi}{4(K^{\prime}+1)}} L_{\frac{1}{2}}(-K^{\prime}),
\end{equation}
\begin{equation}\label{VarRicianDis}
    \textrm{var} \{ \xi^{\prime}_{n} \} \triangleq \rho - \frac{\rho\pi}{4(K^{\prime}+1)} L^{2}_{\frac{1}{2}}(-K^{\prime}),
\end{equation}
with $L_{\frac{1}{2}}(x) =$ $_1F_1(-\frac{1}{2},1,x)$ denoting the confluent hypergeometric function of the first kind.
Then the upper and lower bound of the mean signal power is given by
\begin{equation}
\begin{aligned}
     \bar{S}_{\textrm{ub}} & \triangleq
     \mathbb{E} \{ b_{0}w^2_{\textrm{max}}\left|\Xi\right|^2 \} =
     b_{0}w^2_{\textrm{max}} \left( N^2 \mathbb{E} \{ \xi^{\prime}_n \}^{2} + N \textrm{var} \{ \xi^{\prime}_n \} \right),
\end{aligned}
\end{equation}
\begin{equation}
\begin{aligned}
     \bar{S}_{\textrm{lb}} & \triangleq
     \mathbb{E} \{ b_{0}w^2_{\textrm{min}}\left|\Xi\right|^2 \} =
     b_{0}w^2_{\textrm{min}} \left( N^2 \mathbb{E} \{ \xi^{\prime}_n \}^{2} + N \textrm{var} \{ \xi^{\prime}_n \} \right),
\end{aligned}
\end{equation}
respectively, both of which scale with $N$ in the order of $\mathcal{O}(N^{2})$. 
Therefore, the IRS passive beamforming can effectively compensate for the double pathloss effect and significantly boost the received signal power given sufficiently large $N$.

\subsubsection{IRS random scattering}
A typical scenario for this case is that a non-serving IRS in another cell beamforms towards its own cell and thus generates randomly scattered interference to the typical UE $u$ in the target cell.
In this case, denote $\varepsilon_n \triangleq \zeta_n + \psi_n$ and assume it to be uniformly random in $[0,2\pi)$ due to the random phase $\psi_n$ from the non-serving IRS to the UE.
Similar to the above analysis, we first denote $\Xi^{\prime} \triangleq \sum\nolimits_{n\in\mathcal{N}} w_n \xi^{\prime}_n e^{j\varepsilon_n}$. 
Then, we have the following lemma.
\begin{lemma}\label{CSCGProve}
For i.i.d. $\xi^{\prime}_{n}$ and i.i.d. $\varepsilon_n$, $n\in\mathcal{N}$, by the Lyapunov CLT \cite{CLTtextbook}, $\Xi^{\prime}$ can be approximated by the circularly symmetric complex Gaussian (CSCG) distribution for practically large $N$, i.e.,
\begin{equation}\label{randomscatterDis}
    \Xi^{\prime} \triangleq \sum_{n\in\mathcal{N}} w_n \xi^{\prime}_n e^{j\varepsilon_n} \stackrel{\textrm{approx.}}\sim \textrm{CSCG}\left( 0, \rho \sum\nolimits_{n\in\mathcal{N}} w^2_n \right).
\end{equation}
\end{lemma}
\emph{Proof:}
The variable $\Xi^{\prime}_n \triangleq w_n \xi^{\prime}_n e^{j\varepsilon_n}$ for each element $n$ has zero mean and independent in-phase and quadrature-phase components each with variance $\frac{w_n^2}{2} \rho$, which is proved in Appendix B. Furthermore, since the variables $\Xi^{\prime}_n$, $n \in \mathcal{N}$ are independent, based on the Lyapunov CLT, the independent in-phase and quadrature-phase components can be each approximated by an independent normal distribution with zero mean and variance $\sum\nolimits_{n\in\mathcal{N}} \frac{w_n^2}{2} \rho$ for practically large $N$. Finally, according to the complex random process theory, the distribution of the combined variable $\Xi^{\prime}=\sum\nolimits_{n\in\mathcal{N}} \Xi^{\prime}_n$ is obtained and \textbf{Lemma \ref{CSCGProve}} is thus proved.
\hfill $\blacksquare$

Based on \eqref{Pru}, the interference power from a non-serving IRS to UE $u$ is given by 
\begin{equation}\label{ItfPowerFormula}
    P_{\textrm{r},u} = b_{0} \left| \Xi^{\prime} \right|^2 = b_{0} \left| \sum\nolimits_{n\in\mathcal{N}} w_n \xi^{\prime}_n e^{j\varepsilon_n} \right|^2.
\end{equation}
Similarly, when $w_n$ takes the maximum/minimum value for all $n \in \mathcal{N}$, the mean interference power from the non-serving IRS is thus upper/lower bounded by
\begin{equation}
     \bar{I}_{\textrm{ub}} \triangleq b_{0}
     \mathbb{E} \{ \left|\Xi^{\prime}\right|^2 \}\big|_{w_n=w_\textrm{max},n \in \mathcal{N}} \overset{\mathrm{(a)}}{=}
     b_{0} \rho N w^2_{\textrm{max}},
\end{equation}
\begin{equation}
     \bar{I}_{\textrm{lb}} \triangleq b_{0}
     \mathbb{E} \{ \left|\Xi^{\prime}\right|^2 \}\big|_{w_n=w_\textrm{min},n \in \mathcal{N}} \overset{\mathrm{(a)}}{=} 
     b_{0} \rho N w^2_{\textrm{min}},
\end{equation}
where the equation in (a) is due to \textbf{Lemma 2}. Note that both of them scale with $N$ in the order of $\mathcal{O}(N)$, which is similar to the traditional active beamforming case with $N$ active antennas\cite{QingqingWuTutorial}. 
On the other hand, however, the additional double pathloss effect also applies to the interference power in the IRS-aided scenario, which in turn further mitigates the interference effect.
This is particularly helpful for aerial UEs that suffer from a high probability of strong LoS interference from other cells\cite{IEEEYongZengCellularUAV}, as will be illustrated later in Section \ref{SimulationMultipleCells}.

\subsection{SNR/SINR and Ergodic Throughput}
\subsubsection{Single cell}
The instantaneous received SNR is given by
\begin{equation}
    \gamma_{u} \triangleq P_{\textrm{r},u}/{W},
\end{equation}
where additive white Gaussian noise (AWGN) is assumed with power $W \triangleq N_{0}B$ and power spectrum density $N_0$. Finally, the ergodic throughput within UE $u$'s RB in bits/second/Hz (bps/Hz) can be expressed as
\begin{equation}\label{ergodicThroughput}
    R_{u} \triangleq \mathbb{E} \left\{ \textrm{log}_2(1+\gamma_{u}) \right\}.
\end{equation}

\subsubsection{Multiple cells}
For illustration, consider a cellular system with $Q$ sites shown in Fig. \ref{IRSaidedBSinMultiCells}, each consisting of three cells and thus comprising a total of $C\triangleq 3Q$ cells, denoted by the set $\mathcal{C}$.
In this case, the typical UE $u$ in the target cell 1 suffers from interference from the non-serving IRSs in other cells $\mathcal{C}\setminus \{1\}$.
The aggregate interference power is then given by
\begin{equation}
    I \triangleq \sum\nolimits_{v \in \mathcal{C}\setminus \{1\}} P_{\textrm{r},u,v},
\end{equation}
where the interference power $P_{\textrm{r},u,v}$ from a non-serving cell $v$ can be obtained based on (\ref{ItfPowerFormula}) for the case with random scattering.
On the other hand, the signal power $S$ from the serving cell can be obtained by \eqref{Prphasealign} for the case with passive beamforming.
As a result, the received SINR is given by
\begin{equation}
    \varsigma_{u} \triangleq S/(I + W).
\end{equation}
Finally, the ergodic throughput of UE $u$ can be obtained by replacing $\gamma_{u}$ with $\varsigma_{u}$ in formula (\ref{ergodicThroughput}).





\section{Numerical Results}\label{SectionSimulation}
In this section, we first verify our analytical results in \textbf{Lemma 1} by Monte Carlo (MC) simulations for the $\textrm{I}_n$-UE $u$ channel power gain distribution according to \eqref{hprimtheory} and \eqref{hprimeMC}.
Then, the \textit{3D coverage map} in terms of ergodic throughput distribution in a typical cell is obtained by extensive MC simulations in both single-/multiple-cell scenarios.
In comparison, two benchmark array configuration schemes are also evaluated,
including fixed pattern and 3D beamforming synthesized/enabled by the same number $N = M_{\textrm{Y}}\times M_{\textrm{Z}}$ of active antenna elements, as described in \cite{IEEEYongZengCellularUAV}. In particular, the fixed pattern electrically downtilts its main beam by $10^{\circ}$, while 3D beamforming adopts maximum ratio transmission (MRT). Furthermore, we have incorporated the impact of antenna pattern on the BS-UE channel similarly as in Section \ref{SectionImpactERP}.

In the single-cell scenario, each MC result for a given UE location is obtained by averaging over 25000 samples of BS-UE channel realizations, constituted by 50 instances of LoS/NLoS channel conditions, each with 500 fading realizations.
In the multiple-cell scenario, seven sites are considered as in Fig. \ref{IRSaidedBSinMultiCells}, where each interfering BS-UE channel is generated based on similar procedures in the above.
The following parameters are used if not mentioned otherwise: $f_{\textrm{c}}=2$ GHz,
$d_{\textrm{Y}}=d_{\textrm{Z}}=\lambda/2$, $D=0.559$ m, $N=10 \times 10$, $H_{\textrm{I}}=25$ m, $\mu_{\textrm{Y}}=\mu_{\textrm{Z}}=87^{\circ}$, $A=0.9$, $k=0.5$, $\phi_{\textrm{g}}=3\pi/2$,
$\boldsymbol{U} = \big\{(\theta,\phi) \big| \theta \in \left[ 3\pi/10,\pi/2 \right], \phi \in \left[\pi,2\pi \right] \big\}$,
$P_{\textrm{t}}=10$ dBm, $G_{\textrm{tx}}=8$ dBi, $N_{\textrm{0}}=-174$ dBm/Hz, $B=180$ kHz, $K_{\textrm{min}}=0$ dB, $K_{\textrm{max}}=30$ dB and $C=21$. For ground UEs or aerial UEs, we set $H_{u}=1.5$ m or $120$ m, respectively.
For the IRS-aided BS, two ERPs in the form of \eqref{IRSERPFormula} are considered, with $G_{\textrm{i}}=6$ dBi (i.e., $\cos\theta$) and $G_{\textrm{i}}=9$ dBi (i.e., $\cos^3\theta$), respectively, as is used in \cite{TransShiJinPathlossmodeling} and \cite{TransShiJinPathlossmodelingmmWave}.
For the benchmark schemes, each active antenna element follows the 3GPP ERP as in \cite{IEEEYongZengCellularUAV} with a maximum gain of $G_\textrm{a}=8$ dBi.

Besides, for quantitative comparisons, we also evaluate the average ergodic throughput $\bar R\triangleq \frac{1}{|\mathcal{U}|} \sum_{u\in\mathcal{U}} R_u$, Jain's fairness index\footnote{This index is defined as $J\triangleq \frac{(\sum_{u\in\mathcal{U}} R_u)^2}{|\mathcal{U}| \sum_{u\in\mathcal{U}} R_u^2}$ with $R_u$ denoting the UE throughput in \eqref{ergodicThroughput}, where $J\in(0,1]$ and a higher $J$ represents better fairness.} $J$ and average mean signal power $\bar{ \bar{S}}$ or interference power $\bar{ \bar{I}}$ for ground or aerial UEs in the cell.
The results are summarized in TABLEs \ref{TableSingle} and \ref{TableMultiple}, where 3D BF stands for 3D beamforming while IRScos and IRScos3 denote the IRS-aided BS design with ERP of $\cos\theta$ or $\cos^3\theta$, respectively.

\begin{figure}
	\centering
	\includegraphics[width=0.85\linewidth,  trim=0 0 0 20,clip]{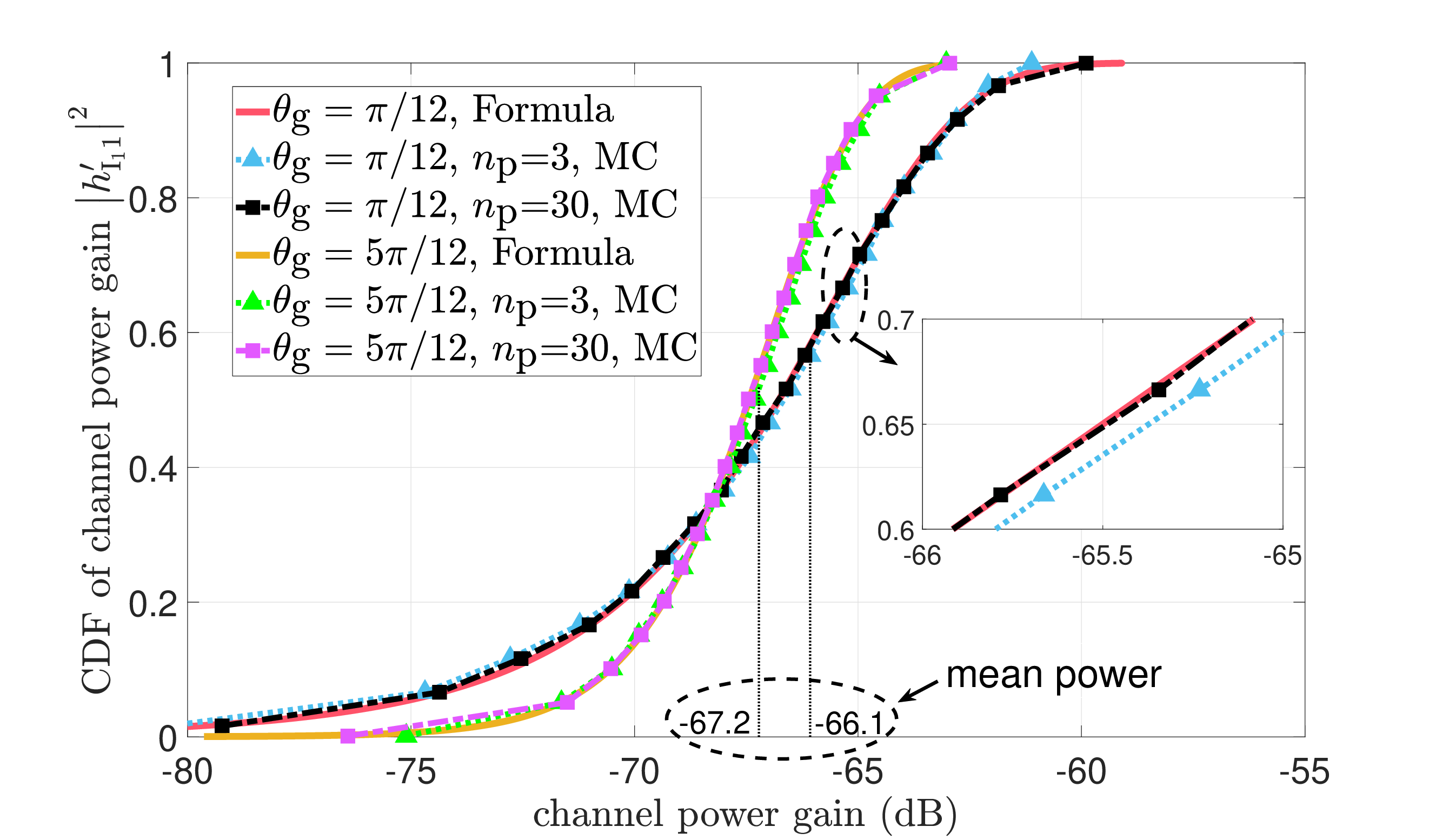}
	\caption{CDF of the channel power gain $\lvert h^{\prime}_{\textrm{I}_1 1}\rvert ^2$, under $\theta_\textrm{m}=\frac{\pi}{12}$, $F_\textrm{i}=\cos \theta$ and different $n_\textrm{p}$ and $\theta_{\textrm{g}}$.\vspace{-1ex}}\label{CDFofIRStoUEchannelpowergain}
\end{figure}

\subsection{Link Level Performance}\label{RicianProveMC}
We first verify the analytical results in \eqref{hprimtheory} and \eqref{hprimeMC} for a given single link (e.g., the link between $\textrm{I}_1$ and a typical ground UE $1$).
Each MC result is obtained by averaging over 3000 randomly generated fading realizations (comprising 10 instances of random path angles, each with 300 random phase realizations), each with a fixed number $n_{\textrm{p}}$ of scattered paths.
Besides, we investigate the \textit{impact of IRS ERP} on multi-path fading channel statistics by comparing two examples of elevation angle distribution in \eqref{PDFEleDistribution} for scattered paths. The first one is with $\theta_{\textrm{g}}=\pi/12$ and $\theta_{\textrm{m}}=\pi/12$ (near the IRS boresight), while the other is with $\theta_{\textrm{g}}=5\pi/12$ and $\theta_{\textrm{m}}=\pi/12$ (far from the IRS boresight). For purpose of illustration, the IRS ERP with $G_{\textrm{i}}=6$ dBi (i.e. $\cos \theta$) is chosen here.

The cumulative distribution function (CDF) of the $\textrm{I}_1$-UE $1$ channel power gain is plotted in Fig. \ref{CDFofIRStoUEchannelpowergain}.
First, it is verified that our analytical results match well with the MC simulation results, especially when the number of scattered paths $n_{\textrm{p}}$ increases.
Second, it is observed that the $\textrm{I}_1$-UE $1$ channel power gain distribution is affected by variations in the scattered path distribution due to the \textit{angle-dependent IRS ERP}.
Specifically, the average channel power gain with $\theta_{\textrm{g}}=\pi/12$ is greater than that with $\theta_{\textrm{g}}=5\pi/12$.
This is because the elevation distribution of scattered paths for $\theta_{\textrm{g}}=\pi/12$ is closer to the boresight of the IRS ERP (i.e., $\theta=0$) than the counterpart of 
$\theta_{\textrm{g}}=5\pi/12$.


\subsection{Single Cell}

\begin{figure}
        \centering
        \begin{subfigure}{1\linewidth}
                \centerline{\includegraphics[width=0.85\linewidth,  trim=0 0 0 0,clip]{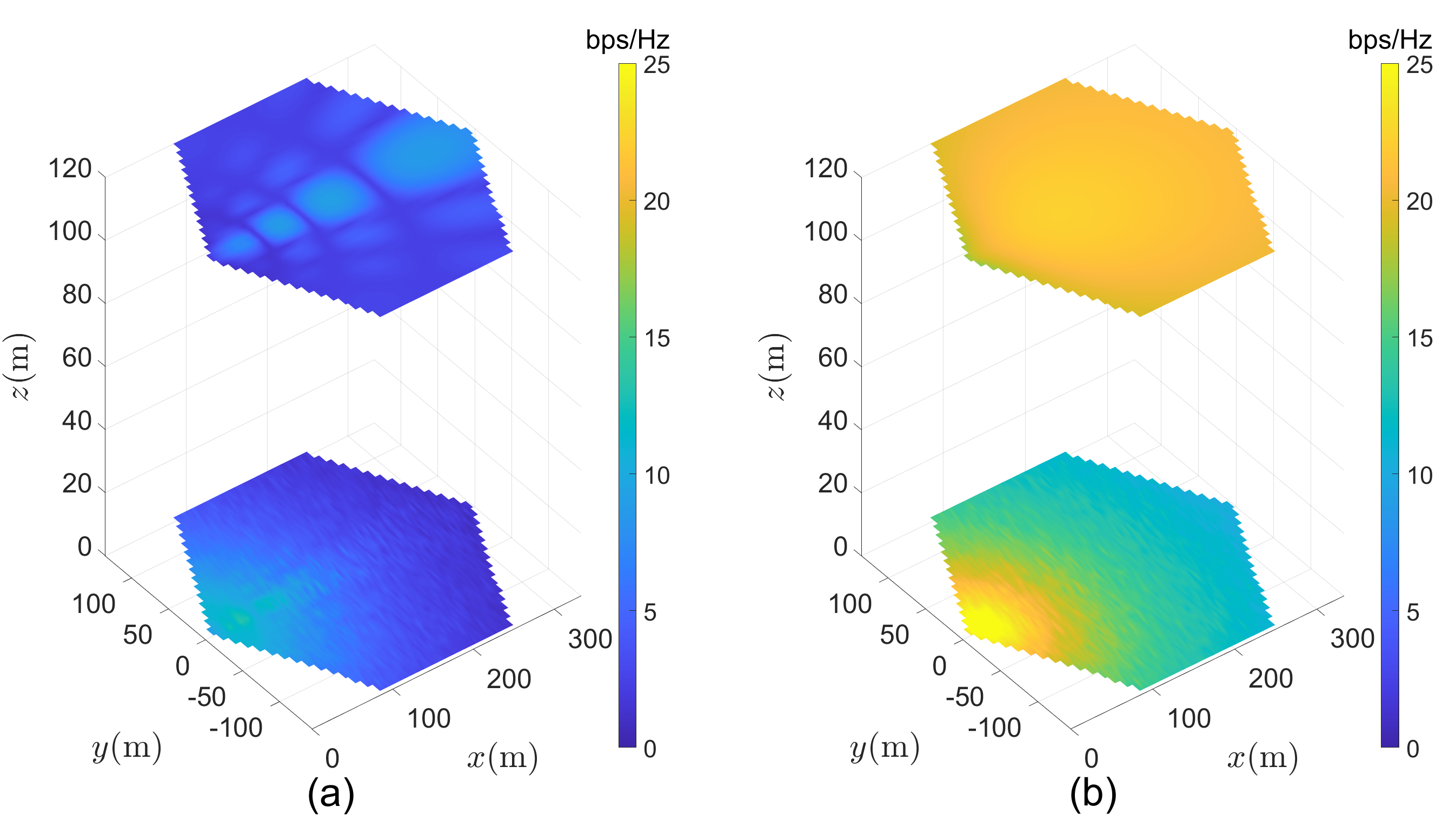}}
                \label{PoutCell0Uplink75}
        \end{subfigure}%
                 
        \begin{subfigure}{1\linewidth}
                \centerline{\includegraphics[width=0.85\linewidth,  trim=0 0 0 0,clip]{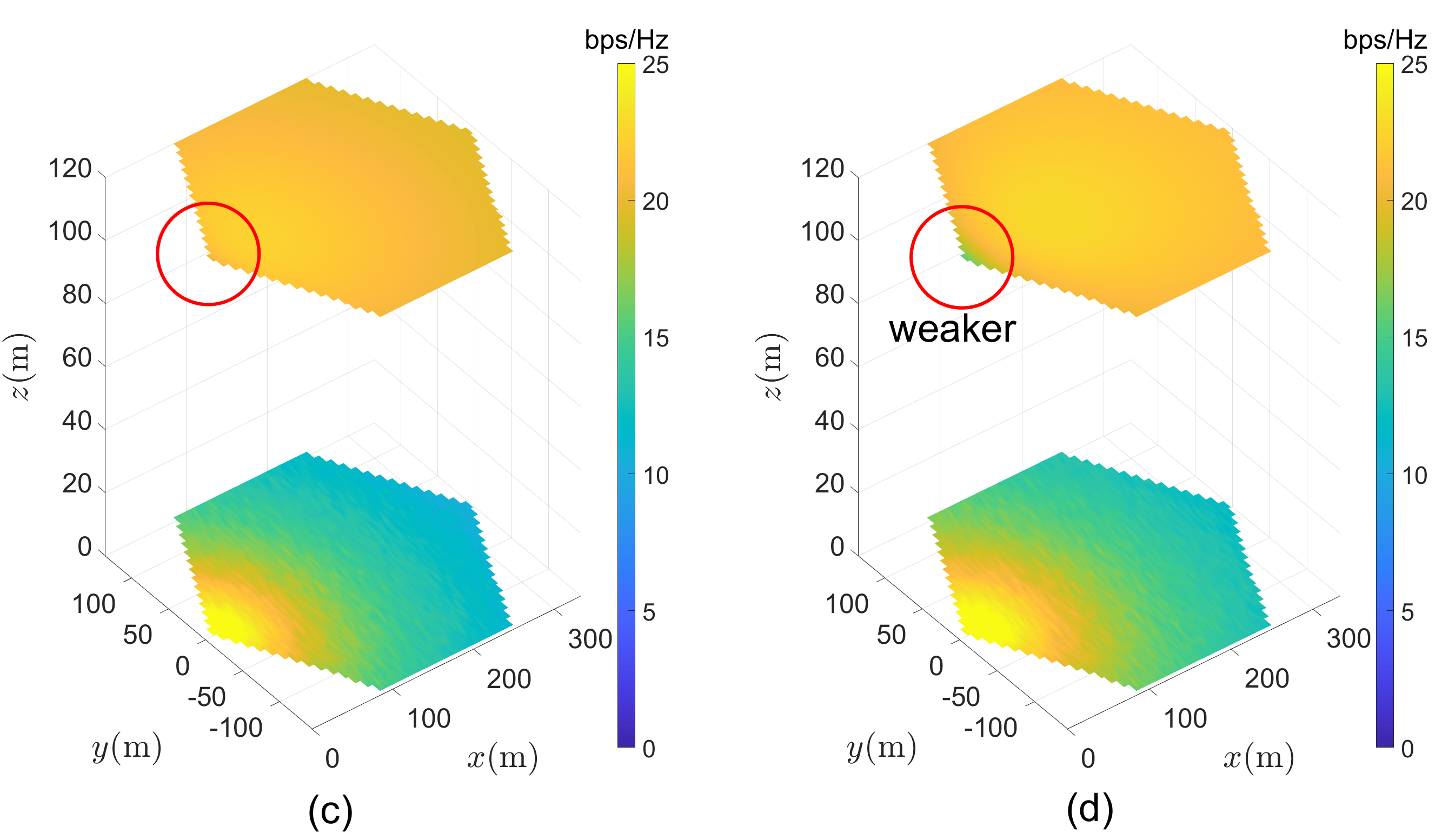}}
                \label{PoutCell0Uplink150}
        \end{subfigure}%
        \caption{Ergodic throughput distribution for ground/aerial UEs in the single-cell scenario, with traditional BS design: (a) fixed pattern (b) 3D beamforming; or IRS-aided BS with ERP: (c) $\cos\theta$ (d) $\cos^3\theta$.\vspace{-2ex}}\label{SingleCellMap}
\end{figure}

The ergodic throughput distribution for ground and aerial UEs in the cell is plotted in Fig. \ref{SingleCellMap}, for the four types of array configurations, respectively.
First, for the fixed pattern case in Fig. \ref{SingleCellMap}(a),
it is shown that the ground coverage performance is better than that in the sky.
This is because the downtilted antenna mainlobe of a traditional BS mainly aims to serve ground UEs, while the aerial UEs can only be served by \textit{weaker antenna sidelobes with possible coverage holes in between}.
Second, for the 3D beamforming case in Fig. \ref{SingleCellMap}(b),
it can be seen that the overall coverage is \textit{much better} and \textit{more balanced} than that of the fixed pattern case, as also suggested by indicators $\bar{R}$ and $J$ in TABLE \ref{TableSingle}.
This is because, by 3D beamforming, the beam angle can be flexibly adjusted towards a target UE to improve its coverage, at the expense of \textit{higher cost and complexity}.
In particular, due to higher LoS probability in the sky, the aerial coverage in turn is better than the ground coverage, thanks to 3D beamforming.

In comparison, for the IRS-aided BS design with ERP of $\cos\theta$ or $\cos^3\theta$, it can be seen from Fig. \ref{SingleCellMap}(c) and (d) as well as Table \ref{TableSingle} that
their average ergodic throughput $\bar{R}$ and fairness $J$ are overall \textit{similar} to that of the 3D beamforming scheme in the single-cell scenario, and yet requiring \textit{much fewer active RF components} (only one active TX with fixed antenna pattern).
The results thus validate the discussions in Section \ref{SectionIRSpassiveBeamRandomScatter} for the impact of IRS passive beamforming.
Interestingly, by comparing Fig. \ref{SingleCellMap}(c) and (d), the overall performance corresponding to the ERP of $\cos^3\theta$ is slightly better than that of the ERP of $\cos\theta$, but is \textit{weaker} at positions right above the BS.
This is because $\cos^3\theta$ has a higher gain but narrower beamwidth, thus the transmitted energy from BS decays faster with angles.




\begin{table}[!htbp]\small
\renewcommand\arraystretch{0.65} 
\centering
\caption{Average ergodic throughput $\bar R$ (bps/Hz), fairness index $J$ and average mean signal power $\bar{\bar S}$ (dB) for ground or aerial UEs in the single-cell scenario.}
\label{TableSingle}
\begin{tabular}{cc|cccc}
   \toprule
   & design & Fixed Pattern & 3D BF & IRScos & IRScos3 \\
   \midrule
   \multirow{4}{*}{\begin{tabular}[x]{@{}c@{}}Ground\end{tabular}}& $\bar R$ & \textbf{4.364} & \textbf{14.951} & \textbf{15.137} & \textbf{16.442} \\
   & $J$ & 0.7218 & 0.9486 & 0.9508 & 0.9613 \\
   & $\bar{\bar S}$ & -112.7 & -79.6 & -80.1 & -77.9 \\
      \midrule
   \multirow{4}{*}{\begin{tabular}[x]{@{}c@{}}Aerial\end{tabular}}& $\bar R$ & \textbf{3.770} & \textbf{20.980} & \textbf{20.905} & \textbf{21.576} \\
   & $J$ & 0.7854 & 0.9986 & 0.9988 & 0.9987 \\
   & $\bar{\bar S}$ & -128.8 & -81.7 & -82.0 & -80.0 \\
   \bottomrule
\end{tabular}
\end{table}

\subsection{Multiple Cells}\label{SimulationMultipleCells}

\begin{figure}
        \centering
        \begin{subfigure}{1\linewidth}
                \centerline{\includegraphics[width=0.85\linewidth,  trim=0 0 0 0,clip]{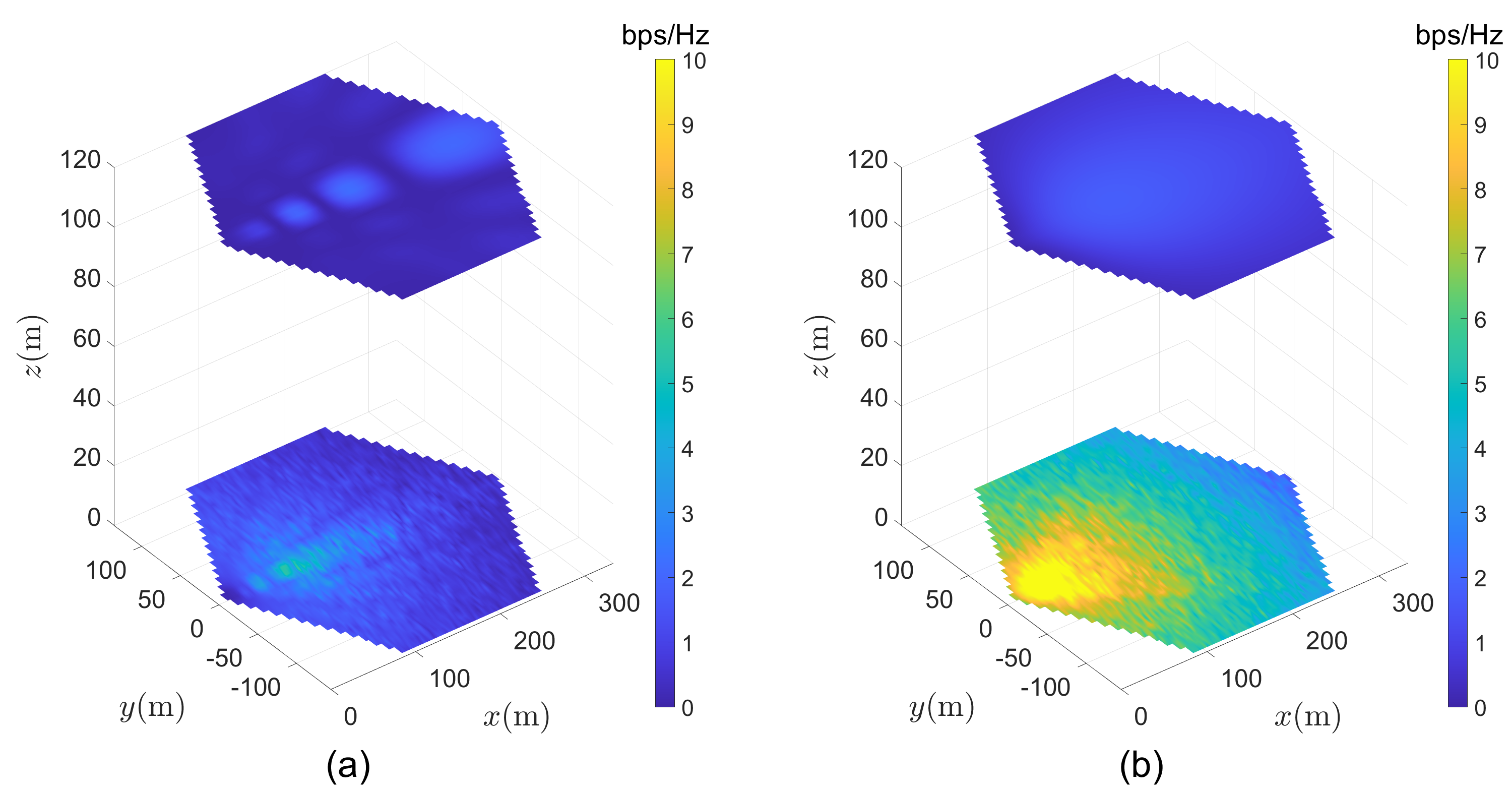}}
                \label{PoutCell0Uplink75}
        \end{subfigure}%
                 
        \begin{subfigure}{1\linewidth}
                \centerline{\includegraphics[width=0.85\linewidth,  trim=0 0 0 0,clip]{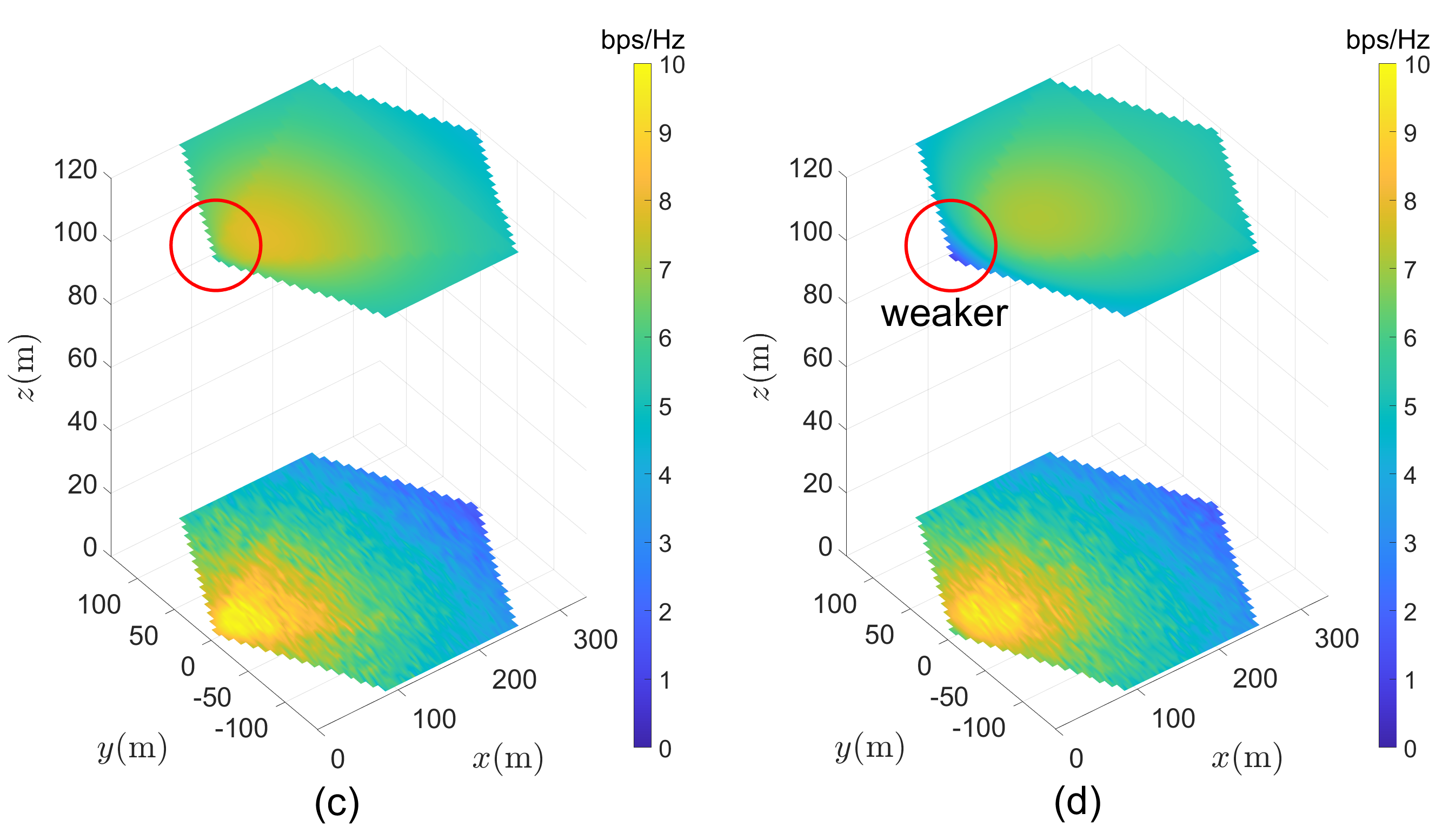}}
                \label{PoutCell0Uplink150}
        \end{subfigure}%
        \caption{Ergodic throughput distribution for ground/aerial UEs in the multiple-cell scenario, with traditional BS design: (a) fixed pattern (b) 3D beamforming; or IRS-aided BS with ERP: (c) $\cos\theta$ (d) $\cos^3\theta$. \vspace{-2ex}}\label{MultiCellMap}
\end{figure}

In the multiple-cell scenario, each IRS-aided sectorized BS follows the same parameters as in the single-cell scenario, except for its boresight that points towards its own cell.
The ergodic throughput distribution in the typical cell 1 is shown in Fig. \ref{MultiCellMap} for the four designs, respectively, whereby the overall coverage performance is degraded compared with that of the single-cell case in Fig. \ref{SingleCellMap} due to inter-cell interference.

Additionally, we have provided numerical validations for the upper/lower bounds of \textit{mean signal or interference power} given by the formulas in Section \ref{SectionIRSpassiveBeamRandomScatter}.
For the example case with the ERP of $\cos\theta$, we have $-85.6 \ \textrm{dB}\leq \bar{\bar{S}} \leq-75.8\ \textrm{dB}, -110.4\ \textrm{dB} \leq \bar{\bar{I}} \leq -100.5\ \textrm{dB}$ for the ground UEs, and $-87.2\ \textrm{dB} \leq \bar{\bar{S}} \leq -77.3\ \textrm{dB}, -106.4 \textrm{dB} \leq \bar{\bar{I}} \leq -96.5\ \textrm{dB}$ for the aerial UEs, both of which are supported by the actual results given by TABLE \ref{TableMultiple}. Similar results are observed for the case with the ERP of $\cos^3 \theta$.

Notably, the aerial coverage by 3D beamforming, though still better than that of fixed pattern, degrades significantly due to the high probability of \textit{strong LoS interference} from other cells.
In comparison, the IRS-aided BS design provides \textit{much better aerial coverage} in the multiple-cell scenario. The underlying reasons are two-folds.
First, based on the analysis in Section \ref{SectionIRSpassiveBeamRandomScatter}, for the link between the serving BS and UE, the $\mathcal{O}( N^2)$ power scaling law of IRS passive beamforming effectively compensates for the double pathloss effect, thus achieving similar mean signal power compared with 3D beamforming in this example, as given in TABLE \ref{TableMultiple}.
On the other hand, for the links between non-serving BSs and the UE, the interference is subjected to the power scaling law of $\mathcal{O}( N)$ in both IRS-aided scheme and active 3D beamforming scheme, whereas the former experiences the \textit{additional double pathloss effect} and hence is much weaker than the latter, as suggested by the indicator $\bar{ \bar{I}}$ in TABLE \ref{TableMultiple}.

\begin{table}[!htbp]\small
\renewcommand\arraystretch{0.65} 
\centering
\caption{Average ergodic throughput $\bar R$ (bps/Hz), fairness index $J$, average mean signal power $\bar{\bar S}$ (dB) and average mean interference power $\bar{\bar I}$ (dB) for ground or aerial UEs in the multiple-cell scenario.}
\label{TableMultiple}
\begin{tabular}{cc|cccc}
   \toprule
   & design & Fixed Pattern & 3D BF & IRScos & IRScos3 \\
   \midrule
   \multirow{4}{*}{
   \begin{tabular}[x]{@{}c@{}}Ground\end{tabular}}& $\bar R$ & 1.207 & 5.852 & 5.522 & 5.466 \\
   & $J$ & 0.7027 & 0.9051 & 0.9020 & 0.9059 \\
   & $\bar{\bar S}$ & -112.8 & -79.6 & -80.1 & -77.9 \\
   & $\bar{\bar I}$ & -120.2 & -99.6 & -104.1 & -101.5 \\
      \midrule
   \multirow{4}{*}{\begin{tabular}[x]{@{}c@{}}\textbf{Aerial}\end{tabular}}& $\bar R$ & 0.370 & \textbf{1.010} & \textbf{6.099} & \textbf{5.757} \\
   & $J$ & 0.3668 & 0.8676 & 0.9820 & 0.9804 \\
   & $\bar{\bar S}$ & -128.8 & -81.7 & -82.0 & -80.0 \\
   & $\bar{\bar I}$ & -125.8 & \textbf{-81.9} & \textbf{-100.0} & \textbf{-97.0} \\
   \bottomrule
\end{tabular}
\end{table}

\section{Conclusion}
This paper proposes an IRS-aided sectorized BS design and analyzes its 3D coverage performance in single-cell and multiple-cell scenarios.
Compared with conventional BSs, the proposed design can enhance the coverage performance in the target area with reduced hardware cost and power consumption.
Numerical results validate our analysis especially the impact of IRS ERP on the coverage performance, 
and demonstrate that our proposed design outperforms the benchmark schemes with fixed BS antenna pattern or active 3D beamforming under the same number of IRS/antenna elements.
Compared with the fixed pattern scheme, it is shown that our proposed design significantly improves the QoS in terms of both achievable throughput and the fairness among UEs.
Compared with the 3D beamforming scheme, the IRS-aided BS achieves similar coverage performance in the single-cell scenario.
Nevertheless, in the multiple-cell scenario, the proposed design provides much better QoS in terms of ergodic throughput compared with active 3D beamforming, for aerial UEs that suffer from strong inter-cell interference.
Finally, wideband phase shifts will be considered in our future work.

\begin{appendices}
\section{The impact of IRS ERP on Rician fading channel}\label{AppenIRSERPonChannel}

When considering isotropic IRS ERP, the Rician faded $\textrm{I}_n$-UE $u$ channel $h_{\textrm{I}_n u}$ can be modeled as
\begin{equation}
    h_{\textrm{I}_n u} = \sqrt{g_u} \left( \sqrt{\frac{K}{K+1}}h_{\textrm{LoS}} + \sqrt{\frac{1}{K+1}}h_{\textrm{NLoS}} \right),
\end{equation}
where $h_{\textrm{LoS}} \triangleq e^{j 2\pi d_u / \lambda}$ denotes the deterministic LoS component with $\left| h_{\textrm{LoS}} \right| = 1$, and $h_{\textrm{NLoS}}$ denotes the NLoS component which can be modeled as a zero-mean and unit-variance CSCG RV.
In practice, the NLoS component $h_{\textrm{NLoS}}$ is the superposition of a certain number of randomly scattered paths, denoted by the set $\mathcal{P}$. Specifically, we have $h_{\textrm{NLoS}} \triangleq \sum\nolimits_{p\in\mathcal{P}}h_{p}$, where $h_p \triangleq a_p e^{-j\varrho}$ corresponds to the channel of the $p^{\textrm{th}}$ path in the set $\mathcal{P}$, with $a_p$ and $\varrho$ denoting its amplitude and phase, respectively.
Based on \cite{tse2005fundamentals}, it is reasonable to assume that the phase for each path is uniformly distributed in $[0,2\pi)$ and that the phases of different paths are independent.
Therefore, by the Lyapunov Central Limit Theorem, $h_{\textrm{NLoS}}$ can reasonably be modeled as a CSCG RV.
By incorporating the non-isotropic IRS ERP, both the LoS and NLoS components will change. Specifically, the LoS component is rewritten as  $h^{\prime}_{\textrm{LoS}} \triangleq \sqrt{G_\textrm{i}F_{\textrm{i}}(\Omega_u)} e^{j 2\pi d_u / \lambda}$, while the NLoS component is $h^{\prime}_{\textrm{NLoS}} \triangleq \sum\nolimits_{p\in\mathcal{P}} \sqrt{G_\textrm{i}F_{\textrm{i}}(\Omega^\textrm{r}_p)} h_p$, where $\Omega^\textrm{r}_p \triangleq (\theta^{\textrm{r}}_p, \phi^{\textrm{r}}_p)$ denotes the AoD of the $p$-th path. As a result, the $\textrm{I}_n$-UE $u$ channel can be rewritten as
\begin{equation}\label{hprimeMC}
    h^{\prime}_{\textrm{I}_n u} = \sqrt{g_u} \left( \sqrt{\frac{K}{K+1}}h^{\prime}_{\textrm{LoS}} + \sqrt{\frac{1}{K+1}}h^{\prime}_{\textrm{NLoS}} \right).
\end{equation}
Note that the IRS ERP has only impact on the amplitude of $h_\textrm{LoS}$ and $h_\textrm{NLoS}$.
As a result, for the NLoS component, each path remains a circular symmetric RV, and hence $h^{\prime}_{\textrm{NLoS}}$ can still be modeled as a CSCG RV, following $\mathcal{CN}\big(0,(K+1)E_{\textrm{NLoS}}\big)$. As a result, $h^{\prime}_{\textrm{I}_n u}$ satisfies the condition of the Rician fading channel, and the Rician K factor can be rewritten as
\begin{equation}
    K^{\prime} \triangleq \frac{\mathbb{E}\{\lvert \frac{K}{K+1} h^{\prime}_{\textrm{LoS}}\rvert ^2\}}{\mathbb{E}\{\lvert \frac{1}{K+1} h^{\prime}_{\textrm{NLoS}}\rvert ^2\}}
    = K \cdot \frac{G_i F_i(\Omega_u)}{(K+1)E_{\textrm{NLoS}}}
    = K \cdot G_K.
\end{equation}
The scale parameters are thus obtained as
\begin{equation}
    \upsilon^{\prime}= \sqrt{ \frac{\mathbb{E}\{(\xi^{\prime}_n)^2\} K^{\prime}}{K^{\prime}+1} } = \sqrt{ \frac{\rho K^{\prime}}{K^{\prime}+1} },
\end{equation}
\begin{equation}
    \sigma^{\prime}= \sqrt{ \frac{\mathbb{E}\{(\xi^{\prime}_n)^2\} }{2(K^{\prime}+1)} } = \sqrt{ \frac{\rho}{2(K^{\prime}+1)} }.
\end{equation}
Therefore, the amplitude $\left| h^{\prime}_{\textrm{I}_n u} \right|$, denoted as $\xi^{\prime}_n$ in \textbf{Lemma 1}, follows the Rician distribution $\textrm{Rice}(\upsilon^{\prime}, \sigma^{\prime})$.



\section{Mean and variance of $\Xi^{\prime}_n$ for the IRS random scattering case}
For the variable $\Xi^{\prime}_n$ in (\ref{randomscatterDis}), denote its amplitude as $V \triangleq w_n \xi^{\prime}_n$ and its phase as $\varphi \triangleq \epsilon_n $. For the IRS random scattering case, the phase $\varphi$ is uniformly random in $[0,2\pi)$, and hence $\Xi^{\prime}_n$ is circularly symmetric with independent in-phase and quadrature-phase components.
In addition,
the amplitude $V$ follows the Rician distribution with the first and second moments given by $w_n \mathbb{E}\{\xi^{\prime}_n\}$ and $w^2_n \mathbb{E}\{ (\xi^{\prime}_n)^2 \}$, respectively. Note that $V$ and $\varphi$ are independent.
Denote $X \triangleq V \cos{\varphi}$ and $Y \triangleq V \sin{\varphi}$ as the in-phase and quadrature-phase components of the $\Xi^{\prime}_n$, respectively.
In the following, we obtain the mean and variance of $X$, while those of $Y$ can be similarly obtained.

First, the first and second moments of $X$ can be expressed as $\mathbb{E}\{X\} = \mathbb{E}\{V \cos{\varphi}\} = \mathbb{E}\{V\}\mathbb{E}\{\cos \varphi\} = 0$ and $ \mathbb{E}\{X^2\} = \mathbb{E}\{ V^2 \cos^{2}{\varphi}\} = \mathbb{E}\{w^2_n (\xi^{\prime}_n)^2 \}\mathbb{E}\{\cos^{2} \varphi\}
    = w^2_n \mathbb{E}\{(\xi^{\prime}_n)^2\} \mathbb{E}\{\frac{1+ \cos(2\varphi)}{2}\} = \frac{w^2_n}{2} \mathbb{E}\{(\xi^{\prime}_n)^2\}$.
Then, the variance of $X$ is given by
\begin{equation}
\begin{aligned}
    \textrm{var}\{X\} &= \mathbb{E}\{X^2\} - \left[\mathbb{E}\{X\}\right]^2 \\
    &= \frac{ w^2_n}{2} \mathbb{E}\{(\xi^{\prime}_n)^2\}= \frac{w^2_n}{2} \rho.
\end{aligned}
\end{equation}
In summary, the variable $\Xi^{\prime}_n \triangleq w_n \xi^{\prime}_n e^{j\varepsilon_n}$ for each element $n$ has zero mean and independent in-phase and quadrature-phase components each with variance $\frac{w_n^2}{2} \rho$.
\end{appendices}

\bibliography{IEEEabrv,bibliography}
\end{document}